# ACCELERATE SINGLE-SHOT DATA ACQUISITIONS USING COMPRESSED SENSING AND FRONSAC IMAGING

*Haifeng Wang, R. Todd Constable and Gigi Galiana*

Department of Diagnostic Radiology, Yale University, New Haven, CT, USA

## ABSTRACT

Nonlinear spatial encoding magnetic (SEM) fields have been studied to complement multichannel RF encoding and accelerate MRI scans. Published schemes include PatLoc, O-Space, Null Space, 4D-RIO, and others, but the large variety of possible approaches to exploiting nonlinear SEMs remains mostly unexplored. Before, we have presented a new approach, Fast ROtary Nonlinear Spatial ACquisition (FRONSAC) imaging, where the nonlinear fields provide a small rotating perturbation to standard linear trajectories. While FRONSAC encoding greatly improves image quality, at the highest accelerations or weakest FRONSAC fields, some undersampling artifacts remain. However, the undersampling artifacts that occur with FRONSAC encoding are relatively incoherent and well suited to the compressed sensing (CS) reconstruction. CS provides a sparsity-promoting convex strategy to reconstruct images from highly undersampled datasets. The work presented here combines the benefits of FRONSAC and CS. Simulations illustrate that this combination can further improve image reconstruction with FRONSAC gradients of low amplitudes and frequencies.

## 1. INTRODUCTION

A number of studies have investigated parallel imaging methods that employ nonlinear SEM fields, including PatLoc imaging [1], O-Space imaging [2-6], Null Space Imaging [7], 4D-RIO [8], etc. The O-space and Null Space approaches in particular illustrate the potential to overcome the limitations of conventional linear spatial encoding and enhance encoding efficiency in the presence of multiple receiver coils. Ultimately, this could allow for greater undersampling thereby shortening scan times [9-10].

Recent research has shown that single shot trajectories using dynamic fields [11-16] may more adequately exploit the potential of nonlinear SEM fields. The work we present here is similar to Ref. [13] and [14], in that it uses time varying nonlinear gradients and is specifically designed to generate compressible waveforms. However, unlike Ref. [13] and [14], where most of the encoding is performed with nonlinear gradients, the nonlinear gradients in this scheme provide only a small perturbation to standard encoding. Adding a rotating nonlinear gradient of modest amplitude provides a flexible strategy to enhance standard undersampled gradient trajectories (i.e. linear trajectories such as EPI, Spiral or Rosette).

We have previously presented this strategy, called FRONSAC imaging [15], which includes linear gradients applied on the three standard linear encoding fields, and rotary or sinusoidal gradient moments, applied on two second-order encoding fields. The low amplitudes of the oscillating gradient waveforms minimally contribute to dB/dt and peripheral nerve stimulation. Images were reconstructed using the Kaczmarz algorithm [2-6]. The simulations results illustrate that FRONSAC imaging can improve image quality in accelerated data acquisitions for arbitrary single-shot trajectories by reducing artifacts due to undersampling. Multi-shot linear trajectories can also incorporate the FRONSAC imaging method.

Compressed sensing (CS) allows sparse signals to be sampled below their conventional Nyquist rate. CS uses a nonlinear procedure to recover the undersampled signal exactly if certain criteria are met [17-25]. Generally, CS theory requires the independent sparse signals to satisfy the restricted isometry property (RIP) condition [17-18]. More specifically, the CS approach generally should satisfy that [19], (i) the image is sparse in a transform domain; (ii) the measurement basis is incoherent with the sparse domain (reached through a sparsifying transform); (iii) a nonlinear reconstruction in the sparse domain enforces sparsity.

The CS approach has been combined with parallel imaging methods with conventional k-space acquisitions [26-30]. But some conventional k-space acquisition schemes, such as Radial and Spiral, can produce more incoherence than standard Cartesian sampling [31-34]. Trajectories can be further improved by adding random perturbations to the conventional sampling trajectories [31-34], but this is often limited by available slew rates. CS recon has successfully been applied to O-Space imaging trajectories to reduce the mean squared error of images at high acceleration, as seen in Ref. [35-36]. However, CS has not been applied to improve the incoherence available in FRONSAC imaging, which has previously only been reconstructed using the standard Kaczmarz algorithm [15].

In this paper, we present a hybrid scheme named CS-FRONSAC, which combines FRONSAC imaging and CS reconstruction. FRONSAC imaging can improve incoherence between the measurement basis and the sparsifying basis, making CS reconstruction more favorable. The results shown below illustrate that the proposed combination can improve image quality compared with either method alone [2-6]. Therefore, CS-FRONSAC can further reduce the mean squared error of images at high acceleration.

## 2. THEORY

### 2.1. Nonlinear Spatial Encoding

It is well known that conventional MRI scanners encode the spatial information of objects using linear magnetic field gradients. But spatial encoding can also be performed with nonlinear magnetic field gradients. Nonlinear gradients may also provide faster gradient field switching within safety limits, spatially-varying resolution, and improved parallel imaging using field shapes that are complementary to radio frequency (RF) receiver coil profiles. Two of the oldest and most widely studied methods are PatLoc [1] and O-Space imaging [2-6].

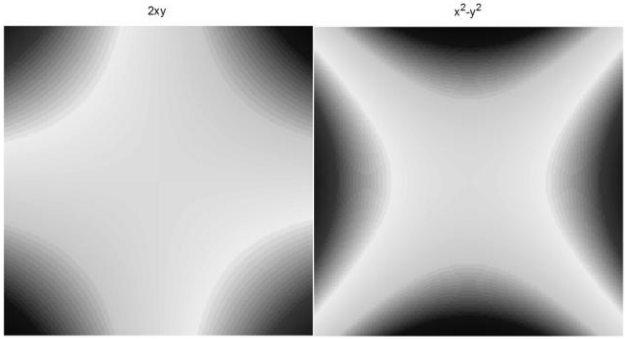

**Fig**.1 Nonlinear second-order gradient field maps of $2xy$ and $x^2-y^2$ fields. Linear combinations of these fields can be used to generate intermediate orientations, and cosine/sine varying amplitudes generate a rotating field.

### 2.2. FRONSAC Imaging

FRONSAC imaging [15] is a flexible scheme to combine linear and second-order SEM fields to improve the image quality of single-shot linear encoding strategies, such as those from heavily undersampled EPI, Spiral, or Rosette trajectories. In this method, standard linear trajectories are applied on the two linear gradient channels, and a rotating gradient is generated by playing sinusoidal gradient waveforms on two second-order encoding fields. Sinusoidal waveforms on the channels $2xy$ and $x^2-y^2$ generate a rotating field similar to those shown in Figure 1. The images are reconstructed using a Kaczmarz algorithm.

If we neglect relaxation effects, the magnetic resonance signal $s_q$ from the q-th RF channel should satisfy the following equation,

$$s_q = \int_\Omega m(\mathbf{x}) C_q(\mathbf{x}) e^{i\Phi(\mathbf{x},t)} d\mathbf{x} \qquad (1)$$

where, $m(\mathbf{x})$ is the magnetization at location $\mathbf{x} = (x, y)$; $C_q(\mathbf{x})$ is the sensitivity profiles of q-th coil, and the integral is over $\Omega$, which is the region of interest; $\Phi(\mathbf{x}, t)$ is the spatially dependent encoding phase. In FRONSAC imaging, the phase factor $\Phi(\mathbf{x}, t)$ in Eq. (1) becomes,

$$\Phi(\mathbf{x},t) = k_x(t)x + k_y(t)y + A \cdot \sin(\omega t) \cdot 2xy \\ + A \cdot \cos(\omega t) \cdot (x^2 + y^2) \qquad (2)$$

where $A$ is the maximum amplitude of the second-order gradient waveforms; $k_x(t)$ and $k_y(t)$ are a vector of gradient moments describing the evolution of each linear gradient field over time. The timing diagram of FRONSAC pulse sequence [15] is shown in Figure 2.

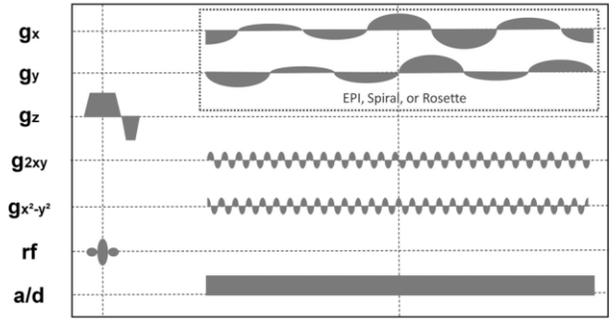

**Fig**.2 Timing diagram of FRONSAC imaging pulse sequence.

Simulations show that FRONSAC imaging can greatly improve many highly undersampled conventional (i.e., linear) imaging schemes, and distinctly eliminate aliasing artifacts allowing for faster imaging speeds. As described in Ref [15], nonlinear gradients spread the sampling in k-space, which allows the sampling of k-space regions that would otherwise be neglected in the undersampled trajectory. The rotation in the FRONSAC gradient then rotates this diffuse sampling function, allowing us to resolve independent values within the sampling functions.

FRONSAC is practical because it provides significant image improvement even at modest nonlinear gradient amplitudes. But even at modest amplitudes, the shortest single shot acquisitions would require high performance gradients to allow rapid switching of the second order gradient fields ($2xy$, $x^2-y^2$). Therefore, techniques that reduce under-sampling artifacts with lower amplitude or frequency FRONSAC gradients are desirable.

### 2.3. CS Reconstruction for FRONSAC

To further decrease the requirements of high performance gradients, we apply CS reconstruction instead of pure Kaczmarz reconstruction. According to previous work on the CS reconstruction for O-Space [35-36], we apply the CS reconstruction with the encoding matrix of FRONSAC imaging instead of the conventional Fourier encoding matrix Φ. With this generalization, the equation may be written as:

$$\mathbf{f} = \arg\min\left\{\|\Phi\mathbf{f}-\mathbf{s}\|_2 + \lambda\|\Psi\mathbf{f}\|_1\right\} \quad (3)$$

In the above equation, **f** is the desired image signal; **s** is the measured signal; $\|\cdot\|_2$ is L-2 norm; $\|\cdot\|_1$ is L-1 norm; Φ is the encoding matrix, i.e. measurement basis or sensing matrix, with its inverse calculated by the Kaczmarz iterative algebraic reconstruction technique, a pseudo-inverse algorithm that converges on the minimum least squares norm solution [2-6]; Ψ is the sparsity transform, such as wavelet, contourlets, finite differences, and so on; λ is the relaxation parameter that controls the convergence properties of the algorithm and is typically set for strongly under-relaxed reconstructions for gradual convergence. The first term imposes data consistency, and the second term encourages sparsity in that domain.

## 3. METHOD

A geometric phantom simulation with resolution, 128×128, is shown for the reference (fully sampled Fourier encoding), FRONSAC, and FRONSAC with CS. The simulations were implemented in MATLAB (MathWorks, Natick, MA, USA) using simulated receiver coil sensitivity profiles from a single uniform birdcage and eight element array [35]. Here three types of single-shot trajectories [15], EPI, Spiral and Rosette, as seen in Figure 3, are considered. EPI trajectories have 8-fold acceleration; the single shot Spiral trajectories have 8 cycles with a constant density; and the Rosette trajectories have single shot trajectories with two frequencies of 98 and 58 Hz if the acquisition time is 25.6ms. The additional FRONSAC gradient is applied with a strength corresponding to A = 790Hz/cm$^2$ and $\omega/2\pi$ = 6.4kHz. The details of CS reconstruction instead of the traditional Kaczmarz algorithm are mentioned in Ref. [35].

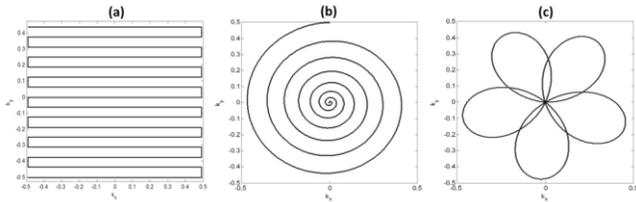

**Fig**.3 Three types of single-shot trajectories, EPI (a), Spiral (b) and Rosette (c), in $k_x$-$k_y$ space

## 4. RESULTS

Simulations are shown in Fig. 4, comparing 3 trajectories based on EPI (a), Spiral (b) and Rosette (c). The 1st row of panel (a) shows reference and CS recon of a standard EPI acquisition. While CS dramatically reduces artifacts, some ghosted features are clearly visible. Below these images in Fig. 4, results are at modest amplitude and frequency (A and ω, corresponding to A = 790 Hz/cm$^2$ and $\omega/2\pi$ = 6.4 kHz).

Even without CS reconstruction, the FRONSAC field perturbation greatly improves image quality, however with CS reconstruction; the result closely resembles the reference image. And other panels to the left and below show reconstructions with weaker amplitude and lower frequency FRONSAC gradients, showing that results are greatly enhanced with CS reconstruction.

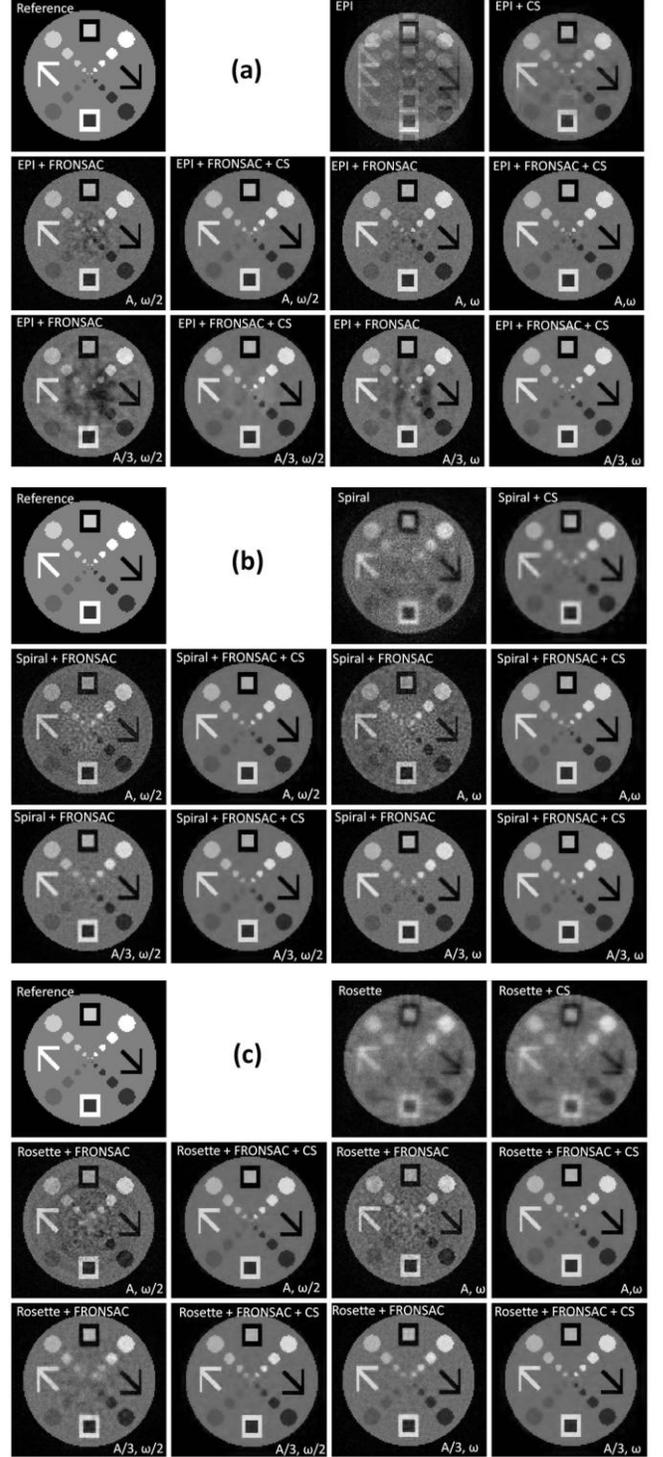

**Fig**.4 Simulated recon results of EPI (a), Spiral (b) and Rosette (c)

Panels (b) and (c) show analogous reconstructions based on Spiral and Rosette trajectories, respectively. Since the gaps in k-space are smaller for these linear trajectories, lower values of A and ω are preferable even without CS reconstruction. However, a similar trend is seen where FRONSAC imaging on its own greatly reduces undersampling artifacts, and the proposed CS-FRONSAC method provides still further improvement, even at lower amplitudes and frequencies.

## 4. CONCLUSIONS

The CS recon can improve image quality when combined with FRONSAC imaging. The findings illustrate that FRONSAC with small encoding gradients, which minimally contribute to PNS or gradient hardware demands, can generate excellent image quality from highly undersampled trajectories.